\documentclass[runningheads,a4paper]{cmmr2023}
\usepackage[style=lncs]{biblatex}
\usepackage{times}
\usepackage{amssymb}
\usepackage{graphicx}
\usepackage{enumitem}
\setlist[itemize]{leftmargin=*}
\setlist[enumerate]{leftmargin=*}
\usepackage{algorithm}
\usepackage{algpseudocode}
\usepackage{tablefootnote}
\usepackage{amsmath}
\usepackage{makecell}
\usepackage{float}
\usepackage{multirow}
\usepackage{booktabs}
\usepackage{amsfonts}
\usepackage{lineno}
\newcommand{\keywords}[1]{\par\addvspace\baselineskip
\noindent\keywordname\enspace\ignorespaces#1}
\begin{document}

\mainmatter  % start of an individual contribution

% first the title is needed
\title{Reconstructing Human Expressiveness in Piano Performances with a Transformer Network}

% a short form should be given in case it is too long for the running head
\titlerunning{Reconstructing Human Expressiveness in Piano Performances with Transformer}

% the name(s) of the author(s) follow(s) next
%
% NB: Chinese authors should write their first names(s) in front of
% their surnames. This ensures that the names appear correctly in
% the running heads and the author index.
%
\author{Jingjing Tang\inst{1} \and Geraint Wiggins\inst{1}\inst{2} \and Gy\"orgy Fazekas\inst{1} \thanks{This work is supported by the UKRI Centre for Doctoral Training in Artificial Intelligence and Music. J.Tang is a research student supported jointly by the China Scholarship Council and Queen Mary University of London. G.~Wiggins received funding from the Flemish Government under the ``Onderzoeksprogramma Artificiële Intelligentie (AI) Vlaanderen''.We would like to thank Lele Liu, Jiawen Huang and the reviewers for their valuable feedback to improve our work.}}
%
% if the names of the authors are too long for the running head, please use the format: AuthorA et al.
\authorrunning{Jingjing Tang et al.}

% the affiliations are given next; don't give your e-mail address
% unless you accept that it will be published
\institute{Center for Digital Music, Queen Mary University of London\\ \and Vrije Universiteit Brussel\\ \email{jingjing.tang@qmul.ac.uk}}

%
% NB: a more complex sample for affiliations and the mapping to the
% corresponding authors can be found in the file "llncs.dem"
% (search for the string "\mainmatter" where a contribution starts).
% "llncs.dem" accompanies the document class "llncs.cls".

\maketitle

\vspace{-3mm}
\begin{abstract}
Capturing intricate and subtle variations in human expressiveness in music performance using computational approaches is challenging. In this paper, we propose a novel approach for reconstructing human expressiveness in piano performance with a multi-layer bi-directional Transformer encoder. To address the needs for large amounts of accurately captured and score-aligned performance data in training neural networks, we use transcribed scores obtained from an existing transcription model to train our model. We integrate pianist identities to control the sampling process and explore the ability of our system to model variations in expressiveness for different pianists. The system is evaluated through statistical analysis of generated expressive performances and a listening test. Overall, the results suggest that our method achieves state-of-the-art in generating human-like piano performances from transcribed scores, while fully and consistently reconstructing human expressiveness poses further challenges. Our codes are released at \url{https://github.com/BetsyTang/RHEPP-Transformer}.

\end{abstract}

\section{Introduction}
An expressive music performance goes beyond playing the notes in the score correctly. Following annotations in music sheets, performers interpret the music with different degrees of expressive control including articulation and dynamics to express emotions and provide an individual rendition of the music, resulting in different performance styles \cite{dai_music_2018}. A common way of rendering expressive performances with computational models is to meaningfully tune the velocity and timing of notes in the score to reconstruct human expressiveness \cite{cancino2018computational}. Generally modelling human expressiveness requires capturing the differences between scores and human performances in expressive features including tempo, timing, dynamics, and so on. Learning the subtle nuances in expression among individual pianists demands the model to learn much smaller perceivable differences within those expressive features.

In recent years, deep learning (DL) models have shown promising results in music generation and representation learning. In particular, the Transformer architecture has gained popularity due to its ability to capture long-range dependencies and contextual information in sequential data. This capability positions the Transformer as a potential solution for modeling performance actions such as adjusting tempo and loudness, and capturing a performer's structural interpretation of music. However, while many studies have successfully applied Transformer architecture to algorithmic music composition \cite{pmlr-v119-choi20b, hsiao2021compound, huang2018music, huang2020pop} and representation learning for symbolic music \cite{chou2021midibert, zeng-etal-2021-musicbert}, few works pay attention to modeling human performance expressiveness independently. In the field of expressive performance rendering (EPR), recent studies have achieved convincing results for the purpose of reconstructing general human expressiveness and controlling style using DL architectures including Recurrent Neural Network \cite{Jeong2019VirtuosoNetAH}, Graph Neural Network \cite{jeong2019graph} and conditional Variational Autoencoder \cite{rhyu2022sketching}. These models require large-scale accurate alignments of well-annotated music scores and performances. However, due to the limited quality and size of the currently available datasets, including the Vienna 4x22 Piano Corpus \cite{goebl2001melody} and ASAP \cite{foscarin2020asap}, these systems still have difficulty dealing with playing techniques such as pedalling and trills, recovering expressiveness overarching longer passages of music, as well as modeling the performance style of individual players.  

In this paper, we propose a novel approach for reconstructing human expressiveness with a multi-layer bi-directional Transformer encoder. Training a Transformer model for this task demands large amounts of accurately recorded and score-aligned performance data, which is not currently readily available. A recently released performance-to-score transcription system \cite{liu2022performance} and the transcribed expressive piano performance dataset ATEPP \cite{zhang2022atepp} allow us to use transcribed scores and performances to train our model. Using transcribed scores in the EPR task can be beneficial when the canonical score is not representative enough. For example, jazz performances rely heavily on  improvisation, making it difficult to align  canonical scores with performances. Even in classical music, ornaments such as trills may not be explicitly notated in canonical scores, which poses problems for the alignment process. Moreover, the reconstruction of human expressiveness from transcribed scores can support research in musical style transfer, particularly when people aim to change a performance by one pianist into the style of another. Considering this, we investigate the ability of our system to model the expressiveness for individual pianists and evaluate it through statistical analysis of the generated performances and a listening test comparing our model to state-of-the-art expressive performance rendering systems.

The rest of this paper is organized as follows: Section \ref{sec:methodology} describes the methodology detailing the dataset, the process of feature extraction and the model architecture. Section \ref{sec:exp} introduces the experiment setting-ups for training our model. Section \ref{sec:evaluation} presents the results of quantitative analysis and the listening test as well as discussions upon the results, and finally, Section \ref{sec:conclusion} concludes the paper.

\section{Methodology}\label{sec:methodology} 
\subsection{Problem Definition}\label{sec:definition}
\begin{figure}[!hbt]
     \centering\vspace{-3mm}\includegraphics[width=0.9\linewidth]{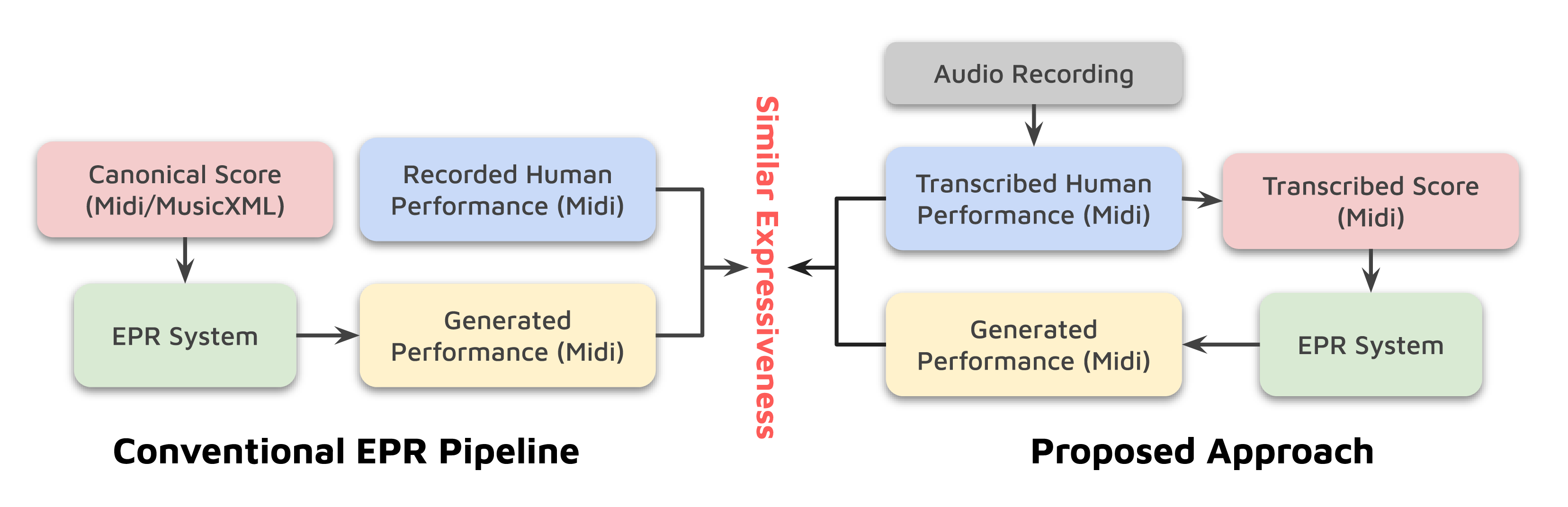}
     \caption{Comparison of the conventional expressive performance rendering (EPR) pipeline with our proposed method}
     \label{fig:definition}
      \vspace*{-10mm}
 \end{figure}

 Expressive performance rendering (EPR) is commonly defined as \textit{the task of generating human-like performances with music sheets as input}. Most existing work \cite{Jeong2019VirtuosoNetAH, jeong2019graph, rhyu2022sketching} proposes systems using recorded performances and canonical scores to solve the problem. All of these systems require alignment between the canonical scores and performances, which is limited in accuracy given the available datasets and alignment algorithms. With the purpose of reconstructing human expressiveness given a composition, we reformulate the task by relaxing the requirement for using conventional music sheets as input, in order to take advantage of the recent performance-to-score transcription algorithms  \cite{liu2022performance} and large transcribed performance datasets  \cite{zhang2022atepp}. We will provide more details about the transcription algorithm and the dataset used in this work in Sections \ref{sec:dataset} and \ref{sec:score_ts}. As shown in Fig.~\ref{fig:definition}, the EPR task, in our definition, is to take the transcribed scores as input and reconstruct human expressiveness by generating expressive performances that are similar to the transcribed human performances.

\subsection{Dataset}\label{sec:dataset}
The recently released ATEPP dataset \cite{zhang2022atepp} provides high-quality transcribed piano performances by world-renowned pianists. According to a listening test conducted by Zhang et al., the transcribed performance MIDIs reliably retain the expressiveness of performers. The dataset includes multiple performances of the same composition by different pianists, allowing comparison in expressiveness among different performers.
\begin{table}[!hbt]
    \caption{Comparison of datasets used in different EPR systems. $^{\star}$NN stands for the number of notes. $^{\dagger}$ \ denotes that the information not provided.}
    \label{tab:dataset_statistics}
    \centering
    \begin{tabular}{lccccc}
    \toprule
\textbf{Systems}&\textbf{Performances}&\textbf{Pianists}&\textbf{Compositions}&\textbf{Composer(s)}&\textbf{Total NN$^{\star}$}\\\hline
VirtuosoNet~\cite{Jeong2019VirtuosoNetAH}&1052&/$^{\dagger}$&226&16&3301K\\
        Sketching-Internal~\cite{rhyu2022sketching}&356&/&34&1&/\\
        Sketching-External~\cite{rhyu2022sketching}&116&/&23&10&/\\
        \hline
         \textbf{Ours}&457&6&36&2&1341K\\
        \hline
    \end{tabular}
    \vspace*{-3mm}
\end{table}
 However, since the ATEPP dataset has a highly skewed distribution of performers, rather than using the whole dataset, we use a subset \cite{syed2023hifi} that balances the number of performances by six pianists:  Alfred Brendel, Claudio Arrau, Daniel Barenboim, Friedrich Gulda, Sviatoslav Richter, and Wilhelm Kempff. Compositions in this subset are mainly composed by Beethoven with only two pieces by Mozart. Each of the compositions corresponds to at least one performance by each pianist. 
 Table ~\ref{tab:dataset_statistics} presents statistics of the subset in comparison with datasets used by other EPR systems.
\vspace{-3mm}
\subsection{Data Processing}
\subsubsection{Score Transcription}\label{sec:score_ts}
Similarly to other EPR systems \cite{Jeong2019VirtuosoNetAH, jeong2019graph, rhyu2022sketching}, our method requires note-to-note alignment between the input score MIDI and the output performance MIDI. Despite the convincing alignment results of the state-of-the-art algorithm proposed by Nakamura et al.\  \cite{Nakamura2017PerformanceED}, the algorithm shows difficulty in dealing with repeated sections as well as trills in classical piano music, which causes unexpected loss of information during the alignment process. Instead of using the original or manually edited scores of the compositions, we obtained the transcribed scores of the performances through a performance-to-score transcription algorithm proposed by Liu et al.\  \cite{liu2022performance}. The transcribed score midi data can be aligned with the performances at the note level without losing any structural generality in the music \cite{WigginsMirandaEtAl93}. 

The transcription algorithm performs rhythm quantisation through a convolutional-recurrent neural network and a beat tracking algorithm to remove expressive variations in timing, velocity, and pedalling. While expressiveness regarding velocity and pedalling is certainly erased through the process, how much expressiveness is remained in timing is implicit and will be discussed further in Section \ref{sec:evaluation}. A further constraint of this algorithm is its inability to retrieve performance directives like dynamics, phrase markings, and beam directions set by the composer. As a result, we were limited to leveraging only the note-related features the algorithm offered.
\vspace{-3mm}
\subsubsection{Data Augmentation}
The transcribed scores are first scaled to the same length as the corresponding performances. We then augment the data by changing the tempo for both performances and the scores. For each pair of performance and score midis, the onset time, offset time and duration of each note are multiplied by a ratio $r_i \in [0.75, 1.25]$. In total, we have each pair augmented by multiplying 10 different ratios that are evenly spaced along the interval grid. 
\begin{table}[!hbt]
    \vspace*{-3mm}
    \caption{Vocabulary size of the tokenized note-level features}
    \label{tab:token_numbers}
    \centering
    \begin{tabular}{cccccc}
    \toprule\textbf{Features}&\textbf{Pitch}&\textbf{Velocity}&\textbf{Duration}&\textbf{Position}&\textbf{Bar}\\\hline
        \textbf{Size}&89&66&4609&1537&518\\
        \hline
    \end{tabular}
    \vspace{-8mm}
\end{table}
\vspace{-2mm}
\subsubsection{Feature Encoding}
Features related to performance expressiveness are extracted and tokenized to reduce the the dimensionality of the input space. Following the tokenisation method, OctupleMIDI, proposed by Zeng et al.\ \cite{zeng-etal-2021-musicbert}, we encode the note-level features including pitch, velocity, duration, bar, and position. 
Table~\ref{tab:token_numbers} shows the vocabulary size of our tokens for each feature. When using OctupleMIDI, the onset time of a note $N_i$ is represented jointly by its bar number $B_i$ and position number $P_i$, where $i = 1, 2, \dots, n$ and $n$ denotes the length of the note sequences. Given that we use a piano music dataset, we  consider only pitches with numbers ranging from 21 to 109. The duration of notes is set to be linearly proportional to the token value $D_i$. All of the midi files have a resolution of 384 ticks per beat, and we default each bar to have 4 beats, resulting in $384 \times 4 = 1536$ different positions per bar. We calculate values of other two note-level performance features which are commonly used for capturing the expressiveness of piano performances \cite{Jeong2019VirtuosoNetAH,ce84fcd7918944e389a7e65b01fe6a9a, rhyu2022sketching} based on the tokens:
\begin{itemize}
    \setlength\itemsep{1em}
    \item \textit{Inter-Onset Interval (IOI)}: the time interval between the onset time (OT) of the note $N_i$ and that of the next note $N_{i+1}$:
    \begin{equation}\label{equ:ioi}
        IOI_{i} =
            \left\{\begin{aligned}
            &OT_{i+1} - OT_{i},\ i=1, 2, \dots, n-1&\\
            &0,\ i=n&
            \end{aligned}\right.
    \end{equation} 
    where  $OT_i = B_i \times 1536 + P_i,\ i = 1, 2, \dots, n$

    \item \textit{Duration Deviation (DD)}: the difference between duration token values of a note in performance midi and score midi
    \begin{equation}\label{equ:dd}
        DD_i = Dp_{i} - Ds_{i}, i = 1,2,\dots, n
    \end{equation}
    where $Dp$ is the duration obtained from the performance midi and $Ds$ is that from the score midi.
\end{itemize}
\subsection{Generation with Transformer Encoder}\label{sec:system}
\subsubsection{Input and Output Features}
 Input and output features are carefully designed to preserve the score content while allowing changes in the performance control of each note. The input features include pitch, velocity, duration, bar, position, and inter-onset interval from the score midis. As for the output, we infer values of three features including velocity, DD, and IOI in the performance midis. Following Eq.~\ref{equ:ioi} and Eq.~\ref{equ:dd}, we can calculate the predicted token values of duration, position, and bar for each note based on DD and IOI. Combined with the predicted token values for velocity, we can construct a performance MIDI file through detokenization.
 
\subsubsection{Model Architecture}\label{sec:model}
\begin{figure}[!hbt]
    \centering
\includegraphics[width=0.55\linewidth]{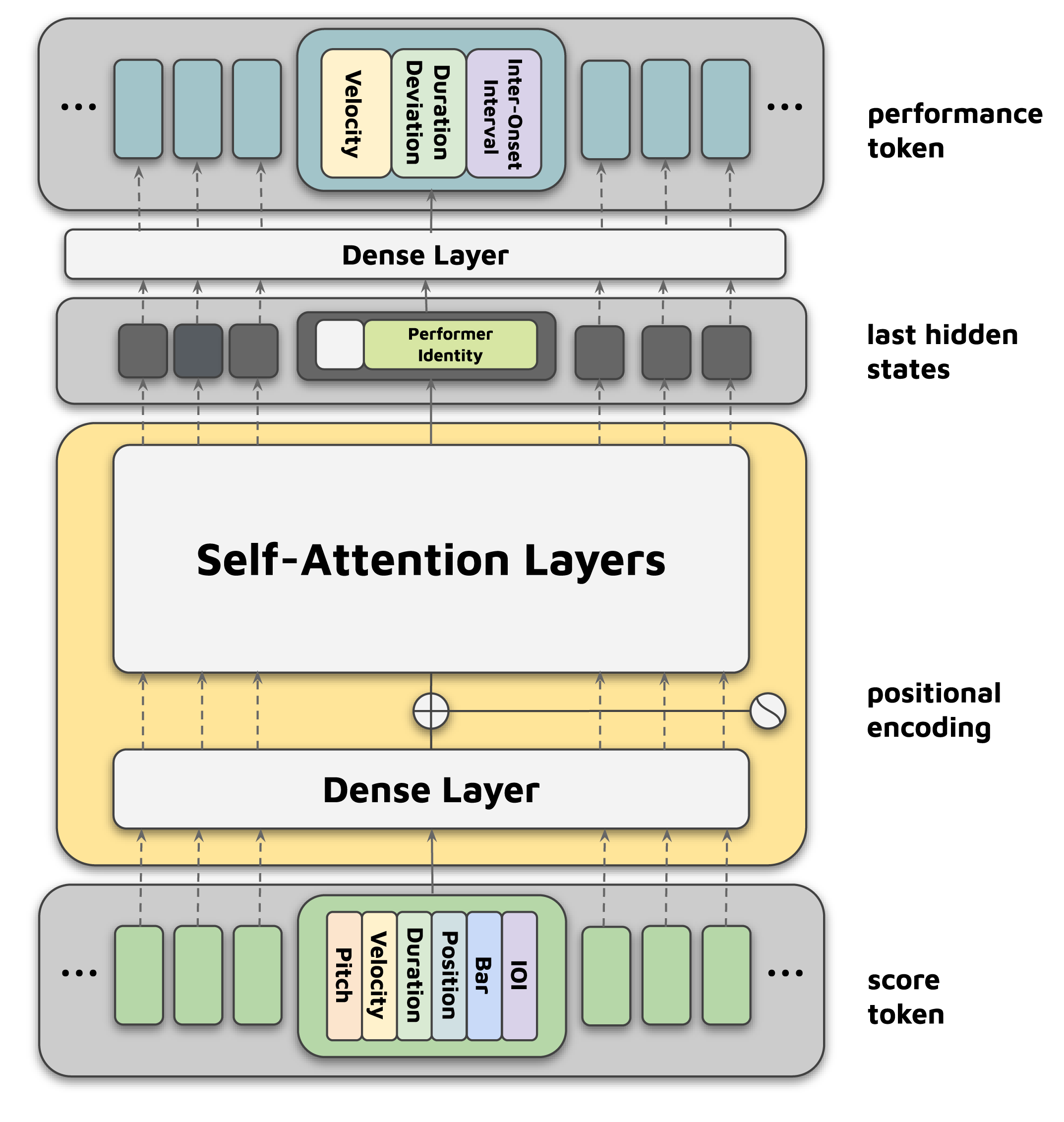}
    \caption{Model architecture of the Transformer encoder}
    \label{fig:model}
    \vspace*{-3mm}
\end{figure}

Inspired by the MidiBert model proposed by Chou et al. \cite{chou2021midibert}, we design a multi-layer bi-directional Transformer encoder with 4 layers of multi-head self-attention where each has 4 heads and a hidden space dimension of 128. The pianist's identity is represented using a one-hot encoding embedding, which is then concatenated to the last hidden state before the final prediction, as shown in Fig ~\ref{fig:model}.
As velocity and timing in music are continuous variables, the interval between two token values is informative in representing the distinction of playing a note. Most transformers trained for music generation \cite{hsiao2021compound, pmlr-v119-choi20b, huang2020pop, huang2018music} take different token values as independent classes which makes this information implicit to the model. Our system instead uses the tokens without creating embeddings, and predicts the token values for different features through regression. In addition, we add activation functions after the inference layer to clamp the predicted values, ensuring that they fall into the ranges of different features.
\vspace{-3mm}
\subsubsection{Loss Design}\label{sec:loss}
%add more citations here
 The losses $\mathcal{L}_v$, $\mathcal{L}_{dd}$, and $\mathcal{L}_{ioi}$ for velocity, DD, and IOI features are calculated respectively, following the loss function defined in Eq.~\ref{equ:loss} which represents the percentage of how much the predicted values $y$ deviated from the target values $\hat{y}$. Masks are created to exclude loss calculation for padded tokens.
 
\begin{equation}\label{equ:loss}
    \mathcal{L}_{feature} = \sum^{n}_{i=0} l(y_i)m_i, 
\end{equation}
where $m_i$ represents the loss mask for the $i$-th note and 
\begin{equation*}
     l(y_i) = \left\{\begin{aligned}
&\frac{|y_i-\hat{y_i}|}{|\hat{y_i}|},\ \text{if } \hat{y_i} \neq 0\\
&\alpha|y-\hat{y_i}|,\ \text{if } \hat{y_i} =0\\
\end{aligned}\right.
\end{equation*}
The parameter $\alpha$ regularizes the loss calculation when the target value is zero and is experimentally set to $0.001$. The total loss is calculated by
\begin{equation}\label{equ:loss_total}
    \mathcal{L}_{total} = w_v\mathcal{L}_v + w_{dd}\mathcal{L}_{dd} + w_{ioi}\mathcal{L}_{ioi}
\end{equation}
where weights are empirically initialized and assigned to each feature loss respectively. 

\subsection{Evaluation}\label{sec:eval-method}
The system is objectively evaluated through validation losses and statistical distributions of expressive parameters in generations, presented in Section ~\ref{sec:eval-stat}. Additionally, we evaluate the perceived expressiveness of generated performances through a subjective listening test. As the aim of EPR task is to generate performances with human-like expressiveness \cite{cancino2018computational}, we assume that the more similar a model's output is to a human performance, the more effectively expressive it is. We recruit participants who have experience in playing musical instruments and who are engaged with classical music, and ask them to rate the presented samples by evaluating how expressive, natural, and human-like they are. The detailed experiment design and conditions and the results of the listening test are presented in Section ~\ref{sec:eval-lt}.

\section{Experimental Setup}\label{sec:exp}
We implement our model based on the PyTorch. We have a 8:1:1 data split in the number of piece and performance, and we cut or pad the token sequences into sequences of 1000 notes before inputting into our transformer. The model is trained with a batch size of 16 sequences for at most 400 epochs, using the Adam optimizer with an initial learning rate of 1e-4 and a weight decay rate of 1e-7. We update the learning rate using the cosine annealing warm restart scheduler \cite{loshchilov2017sgdr} since it has been shown to result in faster convergence during training, compared with other learning rate scheduling strategies. If the validation loss does not improve for 30 consecutive epochs, we stop the training process early. The training converges in 2 days on two RTX A5000 GPUs.

Different vocabulary sizes of expressive features shown in Table~\ref{tab:token_numbers} result in different degrees of complexity when modeling. Consequently, we observed unbalanced decrease in losses and overfitting across learning for different features with constant weights assigned to each feature loss. To balance training and reduce overfitting, we optimize the training process using the GradNorm algorithm proposed by Zhao et al.\ \cite{pmlr-v80-chen18a} to dynamically update weights based on gradients calculated at the end of each training epoch.

\section{Results}\label{sec:evaluation}
\subsection{Quantitative Evaluation}\label{sec:eval-stat}
Quantitative methods for evaluating expressive performance rendering systems are limited. One approach \cite{cancino2018computational} is to calculate the loss for each performance feature. Unlike existing approaches \cite{jeong2019graph, Jeong2019VirtuosoNetAH, rhyu2022sketching} where the features are not tokenised, our system computes the losses using the token values. Based on the feature encoding process and the loss design discussed in Section \ref{sec:methodology}, we estimate the average prediction errors in MIDI quantised velocity value and seconds, shown in Table \ref{tab:loss_values}.
\begin{table}[!hbt]
    \caption{Loss and average prediction error in MIDI velocity value and seconds for note-level expressive features on the test dataset}
    \label{tab:loss_values}
    \centering
    \begin{tabular}{c@{\hskip 0.1in}c@{\hskip 0.1in}c@{\hskip 0.1in}c}
    \toprule    \textbf{Features}&\textbf{Loss}&\textbf{Average Error}\\
    \hline\textbf{Velocity}&0.1267&$\pm$16.2048\\
    \textbf{Duration Deviation}&0.6280&$\pm$0.0473s\\
    \textbf{Inter-Onset Interval}&0.2389&$\pm$0.0183s\\
    \hline
    \end{tabular}
    \vspace*{-3mm}
\end{table}

Although the results are not directly comparable to existing works because of the differences in feature extraction and loss design, they indicate that the transformer model could learn the patterns of expressive variations and reproduce them in the transcribed scores. However, the average errors at the note level in generations are still noticeable to human ears \cite{london2012hearing}, and can affect the perceived expressiveness of the generated music in comparison to human performances.

Since the level of expressiveness regarding timing left in the transcribed scores is implicit as discussed in Section \ref{sec:methodology}, we evaluate the ability of our system to reconstruct the expressiveness for individual pianists through the velocity distributions obtained from kernel density estimation \cite{ce84fcd7918944e389a7e65b01fe6a9a, zhao2021violinist}.
\begin{figure}[!hbt]
    \vspace*{-3mm}
    \centering
\includegraphics[width=\linewidth]{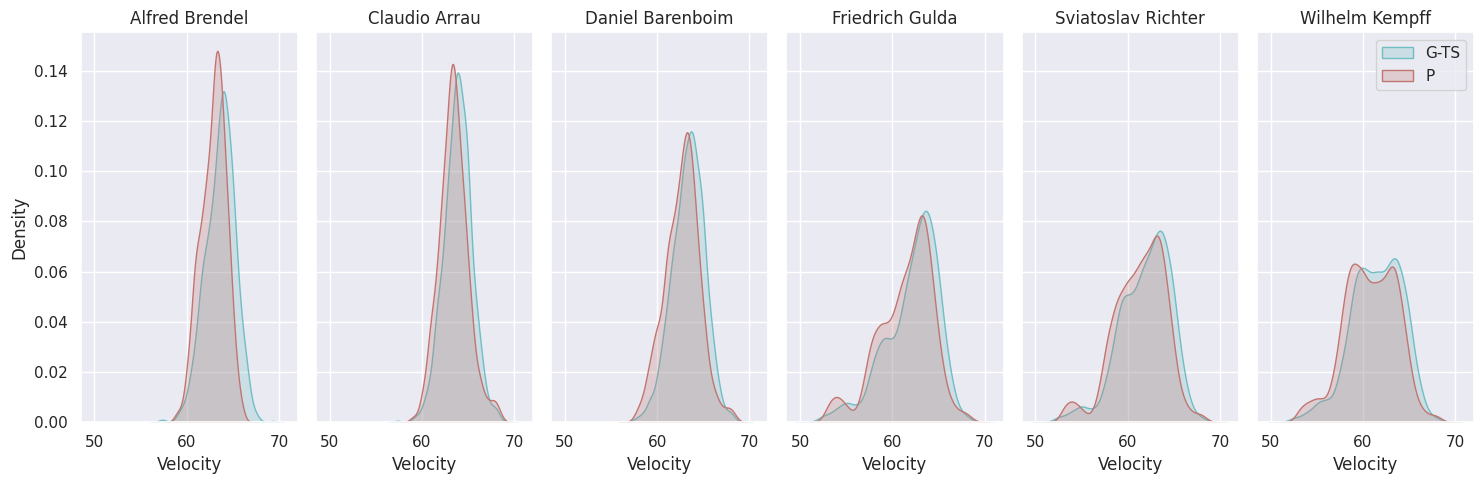}
    \caption{Velocity distributions for the human performances (P) and the our generations (G-TS) on all pieces in the test set, grouped by different pianists.}
    \label{fig:distribution}
     \vspace*{-5mm}
\end{figure}

As shown in Fig.\ \ref{fig:distribution}, velocity distributions for each pianists are distinguishable, indicating different performing styles. However, performance recording environments may have impact on the transcribed velocity values \cite{kong2021high} and contribute to differences of the distributions. The distributions of the generations based on transcribed scores (G-TS) and those of the human performances (P) have a high degree of overlap, providing evidence of learning individual expressiveness through the training.
\vspace{-3mm}
\subsection{Subjective Evaluation}\label{sec:eval-lt}
A listening test was performed to evaluate the perceived expressiveness of our model's output. We recruited 19 people who had some level of music training through email. All participants have learned a musical instrument, while over half of our participants had been engaged with classical music for over 5 years. The participants completed the study anonymously.

The stimuli consisted of four 20s classical piano excerpts detailed in Table ~\ref{tab:compositions}. For each excerpt, the human performance (\textbf{P}) was provided as a reference to be compared with four MIDI renderings: the generation based on the transcribed score (\textbf{G-TS}), the generation by the state-of-the-art VirtuosoNet \cite{Jeong2019VirtuosoNetAH} using the canonical score (\textbf{V}), a direct rendering of the transcribed score (\textbf{TS}), and finally the canonical score (\textbf{S}) without expression. The human performances were transcribed piano performance MIDIs from the ATEPP dataset \cite{zhang2022atepp} and were included as one of the stimuli as well. All the MIDIs were synthesised into audio recordings through GarageBand to ensure consistency in the listening experience. For each piano excerpt, six recordings, the reference plus 5 stimuli, were presented in the test \footnote{Listening samples are provided at\  \url{https://drive.google.com/drive/folders/1nfaZ23vr8xZHlyhTAAppK2hl-aHQPigP?usp=sharing}}. 

Participants were asked to listen to five stimuli, and rate the degree of expressiveness for them on a 100-point scale by comparing each of them with the reference human performance. During the test, we explicitly ask participants to rate based on the expressive differences among the stimulus with more focus on the performance features such as the dynamics and tempo changes rather than the compositional content. We encouraged them to use the full scale, rating the best sample higher than 80 and the worst lower than 20. We adopt the MUSHRA framework \cite{series2014method} to conduct the test using the Go Listen platform \cite{zhang_barry_sun_hines_2021}. 

\begin{table}[!hbt]
    \vspace*{-5mm}
    \caption{Compositions used for the listening test}
    \label{tab:compositions}
    \centering
    \begin{tabular}{l@{\hskip 0.1in}c@{\hskip 0.1in}c}
    \toprule\textbf{Annotation}&\textbf{Composer}&\textbf{Composition}\\
    \hline
    Piece \textbf{A}&Beethoven&\textit{Piano Sonata No. 19 in G Minor, Op. 49 No. 1: II. Rondo (Allegro)}\\
    Piece \textbf{B}&Beethoven&\textit{Piano Sonata No. 7 in D Major, Op. 10 No. 3: III. Menuetto (Allegro)}\\
    Piece \textbf{C}&Haydn&\textit{Piano Sonata in C Major, Hob. XVI:48: II. Rondo (Presto)}\\
    Piece \textbf{D}&Bach&\textit{French Suite No. 5 in G, BWV 816: 7. Gigue}\\
    \hline
    \end{tabular}
    \vspace*{-5mm}
\end{table}

\begin{figure}[!hbt]
\vspace*{-3mm}
    \centering
\includegraphics[width=\linewidth]{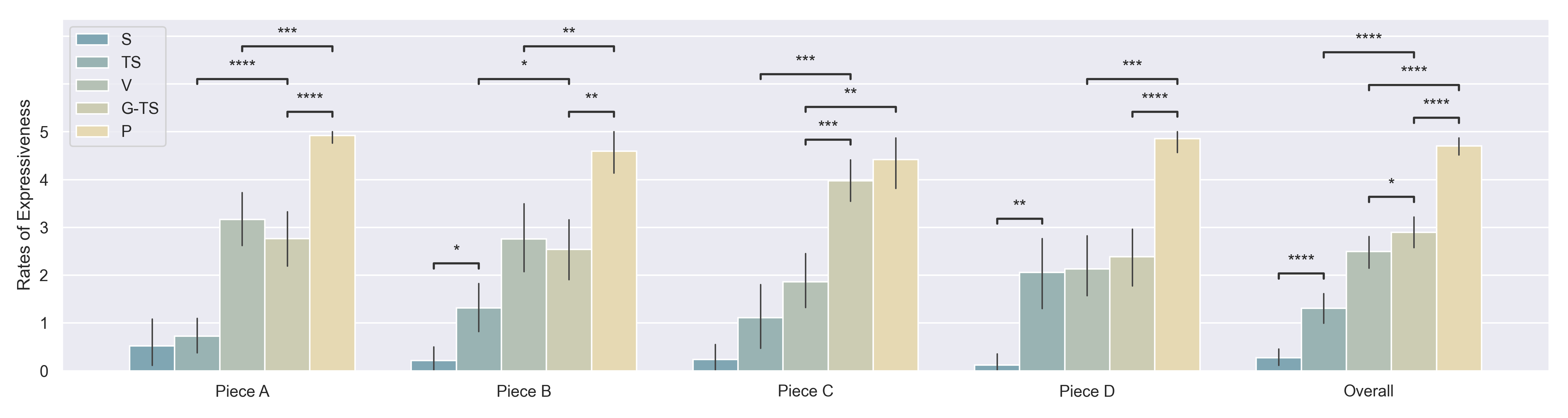}
    \caption{Results of listening test. The mean opinion scores (converted to a 5-point scale) and 95\% confidence intervals are presented for each test piece and the overall results. Wilcoxon signed-rank test are performed to test the significance of the differences. * ($0.01<p<0.05$), ** ($0.001<p<0.01$), *** ($0.0001<p<0.001$), **** ($p<0.0001$)}
    \label{fig:listening test}. 
    \vspace*{-5mm}
\end{figure}

In total, 380 ratings from the 19 listeners were collected. We filtered out raters who could not identify the difference in expressiveness between the anchor (\textbf{S}) and the reference (\textbf{P}). Fig \ref{fig:listening test} shows the mean opinion scores (MOS) and the results of Wilcoxon signed rank test for the differences between: (a) \textbf{TS} versus \textbf{S}, (b) \textbf{G-TS} versus \textbf{V}, (c) \textbf{P} versus \textbf{G-TS}, (d) \textbf{P} versus \textbf{V}, (e) \textbf{G-TS} versus \textbf{TS}.
% Explicit values of MOS and 95\% confidence intervals are presented in Appendix.

According to the results, human performances (\textbf{P}) are significantly different from generations of our model (\textbf{G-TS}) and VirtuosoNet (\textbf{V}) in most situations. The outputs of our model (\textbf{G-TS}) are overall preferred over the performances produced by VirtuosoNet (\textbf{V}) significantly ($0.01 < p < 0.05$), receiving trivially lower (not significant) ratings for piece \textbf{A} and \textbf{B} but higher (significant for \textbf{C} and not significant for \textbf{D}) ratings for the compositions that never appear in the training dataset. Comparing with canonical scores (\textbf{S}), transcribed scores (\textbf{TS}) get significantly higher ratings from listeners. Ratings of the generations by our system (\textbf{G-TS}) are significantly higher than those of the direct audio rendering of transcribed scores (\textbf{TS}) for most pieces except \textbf{D}. 

These results suggest that our system achieves the state-of-the-art and even outperforms the VirtuosoNet \cite{Jeong2019VirtuosoNetAH} in some cases, although neither of the systems can consistently generate the same level of expressiveness as human performances. On the other hand, while the transcribed scores (\textbf{TS}) could have more expressiveness than the canonical scores (\textbf{S}), the generations from the transcribed scores (\textbf{G-TS}) are perceptually more expressive than the transcribed scores (\textbf{TS}) in most cases, indicating the success of reconstructing human expressiveness. The success has also been proven by the overall difference ($0.01 < p < 0.05$) in MOS between our generations (\textbf{G-TS}) and generations from the VirtuosoNet (\textbf{V}).

\vspace{-3mm}
\subsection{Case Study: Comparison in Dynamics and Duration}\label{sec:case}
\begin{figure}[!hbt]
\vspace*{-5mm}
    \centering
\includegraphics[width=0.9\linewidth]{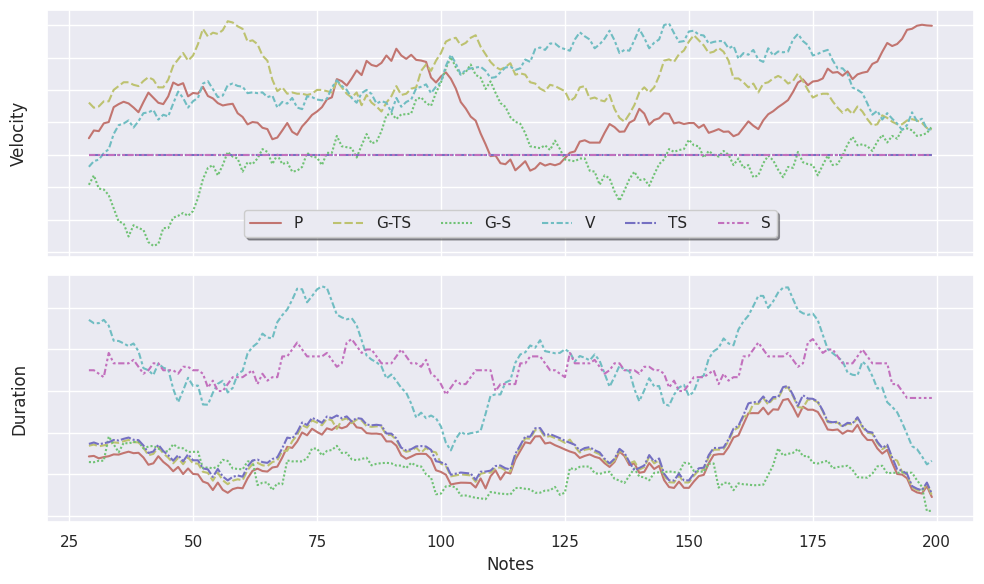}
    \caption{Standardized and smoothed velocity and duration changes across note sequences from \textit{Piano Sonata in C Major, Hob. XVI:48: II. Rondo (Presto)} for enhanced trend comparison. G-S represents the generation of our system based on the canonical scores.}
    \label{fig:example}
    \vspace*{-6mm}
\end{figure}

Building on the promising results of our system in the listening test of Piece \textbf{C}, we conducted a more detailed analysis to compare the expressive variations in dynamics and duration among human performances, system-generated performances, and scores. Specifically, in Fig.\ \ref{fig:example}, we present the fluctuations in velocity and duration across the note sequences. 
Compared with the VirtuosoNet generation (\textbf{V}), the generation of our system (\textbf{G-TS}) could capture both short-term and long-term velocity variations better. Even when inputting the unseen canonical score, the generation of our system (\textbf{G-S}) outperforms the other model in terms of reconstructing velocity variations. Meanwhile, the strong similarity between duration changes in the human performance (\textbf{P}) and transcribed score (\textbf{TS}) suggest that the transcription algorithm \cite{liu2022performance} alters the timing information of the notes cautiously with only limited modification of the duration. Therefore, the reconstruction of the expressive variations in timing through our system could be restricted. The limitation is also demonstrated by the duration changes of our system's generation based on the canonical score (\textbf{G-S}).
\vspace{-3mm}
\section{Conclusion}\label{sec:conclusion}
\vspace{-2mm}
This paper presents a novel method for reconstructing human expressiveness in classical piano performances. Our expressive performance rendering system consists of a Transformer encoder trained on transcribed scores and performances. The quantitative evaluation and listening test show that the proposed method succeed in generating human-like expressive variations, especially for dynamics. Moreover, our method could be used for modeling the differences in expressiveness among individual pianists. 

In future work, we will train our system with a mixture of the canonical scores and transcribed scores to create a more robust system. We will further improve the capacity of our system on modeling individual performance styles possibly through contrastive learning. In addition, we will consider a separate system to model pedalling techniques in performances or try to integrate the pedalling information into the current feature encoding.

% \section*{Appendix}\label{sec:appendix}
%  \vspace*{-8mm}
% \begin{table*}[!hbt]
%     \centering
%     \begin{tabular}{c@{\hskip 0.1in}c@{\hskip 0.1in}c@{\hskip 0.1in}c@{\hskip 0.1in}c@{\hskip 0.1in}c}
%     \toprule
%     &\textbf{Piece A}&\textbf{Piece B}& \textbf{Piece C}&\textbf{Piece D}&\textbf{Overall}\\
%     \hline
%     \textbf{P} (Reference)&4.92$\pm$0.17&4.59$\pm$0.47&4.42$\pm$0.61&4.85$\pm$0.31&4.76$\pm$0.16\\
%     \hline
%     \textbf{S}&0.52$\pm$0.54&0.22$\pm$0.31&0.24$\pm$0.32&0.12$\pm$0.25&0.29$\pm$0.17\\
%     \textbf{TS}&0.72$\pm$0.41&1.32$\pm$0.57&1.11$\pm$0.78&2.06$\pm$0.82&1.27$\pm$0.27\\
%     \textbf{V}&\textbf{3.16$\pm$0.63}&\textbf{2.76$\pm$0.81}&1.86$\pm$0.65&2.13$\pm$0.70&2.55$\pm$0.29\\
%     \textbf{G-TS}&2.76$\pm$0.65&2.54$\pm$0.71&\textbf{3.98$\pm$0.50}&\textbf{2.38$\pm$0.68}&\textbf{2.70$\pm$0.30}\\
%     \hline
%     \end{tabular}
%     \vspace*{2mm}
%     \caption{The mean opinion scores (MOS) and 95\% confidence intervals of the listening test}
%     \label{tab:mos}
%     \vspace*{-10mm}
% \end{table*}
\vspace{-3mm}
\printbibliography
\end{document}